\def\maketitle{\par
 \begingroup
 \def\thefootnote{\fnsymbol{footnote}}
 \def\@makefnmark{\mbox{$^\@thefnmark$}}
 \@maketitle
 \@thanks
 \endgroup
 \setcounter{footnote}{0}
 \let\maketitle\relax
 \let\@maketitle\relax
 \gdef\@thanks{}\gdef\@author{}\gdef\@title{}\let\thanks\relax}
\def\@maketitle{\vspace*{0.9cm}
{\hsize\textwidth
 \linewidth\hsize \centering
 {\normalsize \bf \@title \par}
\vskip 0.3cm  {\normalsize  \@author \par}}}

\def\thefootnote{\mbox{\noindent$\fnsymbol{footnote}$}}
    \long\def\@makefntext#1{\noindent$^{\@thefnmark}$#1}

\def\section{\@startsection{section}{1}{\z@}{1.5ex plus 0.5ex minus
   1.2ex}{1.3ex plus .1ex}{\normalsize\bf}}
\def\subsection{\@startsection{subsection}{2}{\z@}
   {1.5ex plus 0.5ex minus 1.2ex}{1.3ex plus .1ex}{\normalsize\em}}

\def\@sect#1#2#3#4#5#6[#7]#8{\ifnum #2>\c@secnumdepth
     \def\@svsec{}\else
     \refstepcounter{#1}\edef\@svsec{\ifnum #2=1 \@sectname\fi
        \csname the#1\endcsname.\hskip 1em }\fi
     \@tempskipa #5\relax
      \ifdim \@tempskipa>\z@
        \begingroup #6\relax
          \@hangfrom{\hskip #3\relax\@svsec}
          {\interlinepenalty \@M #8\par}
        \endgroup
       \csname #1mark\endcsname{#7}\addcontentsline
         {toc}{#1}{\ifnum #2>\c@secnumdepth \else
                      \protect\numberline{\csname the#1\endcsname}\fi
                    #7}\else
        \def\@svsechd{#6\hskip #3\@svsec #8\csname #1mark\endcsname
                   {#7}\addcontentsline
                     {toc}{#1}{\ifnum #2>\c@secnumdepth \else
                     \protect\numberline{\csname the#1\endcsname}\fi
                   #7}}\fi
     \@xsect{#5}}

\def\@sectname{}

\def\thebibliography#1{\section*{{{\normalsize
\bf References }
\rule{0pt}{0pt}}\@mkboth
  {REFERENCES}{REFERENCES}}\list
  {{\arabic{enumi}.}}{\settowidth\labelwidth{{#1}}%
    \leftmargin\labelwidth  \frenchspacing
    \advance\leftmargin\labelsep
    \itemsep=-0.2cm
    \usecounter{enumi}}
    \def\newblock{\hskip .11em plus .33em minus -.07em}
    \sloppy
    \sfcode`\.=1000\relax}

\def\@cite#1#2{\unskip\nobreak\relax
    \def\@tempa{$\m@th^{\hbox{\the\scriptfont0 #1}}$}%
    \futurelet\@tempc\@citexx}
\def\@citexx{\ifx.\@tempc\let\@tempd=\@citepunct\else
    \ifx,\@tempc\let\@tempd=\@citepunct\else
    \let\@tempd=\@tempa\fi\fi\@tempd}
\def\@citepunct{\@tempc\edef\@sf{
\spacefactor=\the\spacefactor\relax}\@tempa
    \@sf\@gobble}

\def\citenum#1{{\def\@cite##1##2{##1}\cite{#1}}}
\def\citea#1{\@cite{#1}{}}

\newcount\@tempcntc
\def\@citex[#1]#2{\if@filesw\immediate\write\@auxout{
      \string\citation{#2}}\fi
  \@tempcnta\z@\@tempcntb\m@ne\def\@citea{}\@cite{
      \@for\@citeb:=#2\do
    {\@ifundefined
       {b@\@citeb}{\@citeo\@tempcntb\m@ne
       \@citea\def\@citea{,}{\bf ?}\@warning
       {Citation `\@citeb' on page \thepage \space undefined}}%
    {\setbox\z@\hbox{\global\@tempcntc0\csname b@\@citeb
       \endcsname\relax}%
     \ifnum\@tempcntc=\z@ \@citeo\@tempcntb\m@ne
       \@citea\def\@citea{,}\hbox{\csname b@\@citeb\endcsname}%
     \else
      \advance\@tempcntb\@ne
      \ifnum\@tempcntb=\@tempcntc
      \else\advance\@tempcntb\m@ne\@citeo
      \@tempcnta\@tempcntc\@tempcntb\@tempcntc\fi\fi}}\@citeo}{#1}}
\def\@citeo{\ifnum\@tempcnta>\@tempcntb\else\@citea\def\@citea{,}%
  \ifnum\@tempcnta=\@tempcntb\the\@tempcnta\else
   {\advance\@tempcnta\@ne\ifnum\@tempcnta=\@tempcntb
       \else \def\@citea{--}\fi
  \advance\@tempcnta\m@ne\the\@tempcnta\@citea\the\@tempcntb}\fi\fi}

\def\abstract{\if@twocolumn
\section*{Abstract}         
\else \small
\begin{center}
{ABSTRACT\vspace{-.5em}\vspace{0pt}}
\end{center}
\quotation
\fi}
\def\endabstract{\if@twocolumn\else\endquotation\fi}

\def\fnum@figure{Fig. \thefigure}

\long\def\@makecaption#1#2{
   \vskip 10pt
   \setbox\@tempboxa\hbox{\small #1. #2}
   \ifdim \wd\@tempboxa >\hsize    
      \small #1. #2\par            
   \else                           
      \hbox to\hsize{\hfil\box\@tempboxa\hfil}
   \fi}

\documentstyle[12pt,worldsci]{article}
\newcommand{\NPB}[1]{{\it Nucl. Phys.}\ {\bf B{#1}}}
\newcommand{\PLB}[1]{{\it Phys. Lett.}\ {\bf B{#1}}}
\newcommand{\PRD}[1]{{\it Phys. Rev.}\ {\bf D{#1}}}
\newcommand{\PRL}[1]{{\it Phys. Rev. Lett.}\ {\bf #1}}
\newcommand{\hc}{ {\rm h.c.} }
\newcommand{\ME}{ M_{ETC} }
\newcommand{\gE}{ g_{ETC} }
\newcommand{\bb}{ {b{\bar b}} }
\newcommand{\Zbb}{ {Zb{\bar b}} }
\newcommand{\sw}{ s_\theta }
\newcommand{\cw}{ c_\theta }
\newcommand{\esc}{ {e\over\sw\cw} }
\newcommand{\half}{{1 \over 2}}
\newcommand{\gae}{\raise -.15truecm\vbox{
  \hbox{$\; >\;$}\vskip -.17truecm
\hbox{$\; \sim\;$}}}

\begin{document}

\title{{\bf EXTENDED TECHNICOLOR CORRECTIONS TO \\
THE $Z b\bar b$ VERTEX}
\thanks{Talk presented at Beyond the Standard Model
III, June 22-24, 1992, Ottawa, CA. Preprint number HUTP-92/A024.}
}
\author{ELIZABETH H. SIMMONS\\
{\em Lyman Laboratory of Physics, Harvard University,
Cambridge MA 02138, USA}\\
\vspace{0.3cm}
and\\
\vspace*{0.3cm}
R. SEKHAR CHIVUKULA and STEPHEN B. SELIPSKY\\
{\em Department of Physics, Boston University, 590 Commonwealth
Avenue\\ Boston, MA  02215, USA}}

\maketitle
\setlength{\baselineskip}{2.6ex}

\begin{center}
\parbox{13.0cm}
{\begin{center} ABSTRACT \end{center}
{\small \hspace*{0.3cm}
Extended technicolor theories generate potentially large corrections to the
$\Zbb$ vertex.  These can be observed in current experiments at LEP.}}
\end{center}

The origin of the diverse masses and mixings of the quarks and leptons
remains a mystery; most puzzling of all is the origin of the top quark's
large mass.  In technicolor models\cite{1}, the large top mass is thought
to arise from extended technicolor\cite{2} (ETC) dynamics at relatively low
energy scales\footnote{So long as no additional light  scalars couple to
ordinary and techni- fermions\cite{3,4}.}.  Since the magnitude of the KM
matrix element $|V_{tb}|$ is very nearly one, $SU(2)_W$ gauge invariance
ensures that the ETC dynamics responsible for generating the top mass also
couples to the  left-handed component of the bottom quark.  We wish to
point out that this  means the same dynamics will produce potentially large
``non-oblique''\cite{5}  effects at the $\Zbb$ vertex\footnote{This was
first discussed in ref. 6.}.  In particular, if $m_t \gae 100$ GeV and no
effect is visible with data currently being obtained at LEP, theories in
which the ETC and weak interactions commute ({\it i.e.} in which the ETC
gauge bosons are $SU(2)_W$ singlets) can be ruled out, with the same
confidence as models with excessive flavor changing neutral currents.

Consider a model in which the top mass is generated by the exchange of
a weak-singlet ETC gauge boson of mass $M_{ETC}$.  That gauge boson
will carry technicolor and will couple with strength $\gE$ to the current
\begin{equation}
  \xi {\bar\psi^i}_L \gamma^\mu T_L^{iw} +
   {1\over \xi} {\bar t_R} \gamma^\mu U_R^w ,
\label{tmasscur}
\end{equation}
where $\psi_L = (t,b)_L$ is the left-handed weak doublet of
third-generation quarks, $T_L = (U,D)_L$ is a left-handed weak doublet of
technifermions, and $U_R$ is a corresponding right-handed weak-singlet
technifermion.  The indices $i$ and $w$ are weak and  technicolor indices,
while the constant $\xi$ is an ETC Clebsch which is  expected to be of
order one.  At energies lower than $\ME$, ETC  gauge boson exchange may be
approximated by local four-fermion operators.   For example, we can
represent the ETC boson exchange which produces the top quark mass by
writing down the local four-fermion operator coupling  the left- and
right-handed pieces of the current in Eq. (\ref{tmasscur})
\begin{equation}
   -{\gE^2 \over \ME^2}  \left({\bar\psi}_L^i \gamma^\mu
T_L^{iw}\right) \left( {\bar U^w}_R \gamma_\mu t_R \right) + \hc\ .
\label{topff}
\end{equation}
When this is Fierzed into a product of technicolor singlet densities
 \begin{equation}
 2 {\gE^2\over\ME^2} \left( {\bar\psi^i}_L
   t_R \right) \left( {\bar U}_R T^i_L \right) + \hc\ .
\label{topfierz}
\end{equation}
it may be seen to give rise to the top mass.

We can use the rules of naive dimensional analysis\cite{7} to estimate the
size of the top quark mass generated by Eq. (\ref{topfierz}) .   Assuming
(for simplicity) that there is only one doublet of technifermions, that the
strong technicolor interactions respect an $SU(2)_L \times SU(2)_R$ chiral
symmetry, and therefore that the technicolor $F$ constant (analogous to
$f_\pi$ in QCD) is $v\approx 250$ GeV we find
\begin{equation}
   m_t\ = {\gE^2 \over \ME^2}
   \langle{\bar U}U\rangle\ \approx\ {\gE^2 \over \ME^2} (4\pi v^3)\ .
\label{topmass}
\end{equation}
Equivalently, we may express this as an equation determining
the scale of the ETC dynamics responsible
for generating the top mass
\begin{equation}
 \ME \approx 1.4\ {\rm TeV} \cdot
   \gE \left({100\ {\rm GeV} \over m_t} \right)^\half\ .
\label{etcmass}
\end{equation}
In the absence of fine tuning\cite{3} and as long as $\gE^2 v^2/\ME^2
< 1$ (or, equivalently, when $m_t/4\pi v$ is small), the ETC interactions
may be treated as a small perturbation on the technicolor dynamics and
these estimates are self-consistent.  Note that the rules of naive
dimensional analysis do not require that $\ME$ be large, only that
$m_t/4\pi v$ be small.

These dimensional estimates are typically modified in ``walking
technicolor'' models\cite{8} where there is an enhancement of operators of
the form (\ref{topfierz}) due to a large anomalous dimension of the
technifermion mass operator.  The enhancement is important for the ETC
interactions responsible for light fermion masses (for which $\ME$ must be
quite high), but will not be numerically significant in the case of the top
quark because the ETC scale (\ref{etcmass}) associated with the top quark
is so low.  Hence, the results in ``walking'' theories are expected to be
similar to those presented below.

To see how the ETC boson responsible for the top quark mass affects the
$\Zbb$ vertex, we now consider the four-fermion
operator\footnote{Ref. 9 lists four-fermion operators arising from ETC
boson exchange, with emphasis on model-dependent and potentially dangerous
contributions to $\delta\rho$.}
arising purely from the left-handed part of the current (\ref{tmasscur})
\begin{equation}
  -\xi^2 {\gE^2\over\ME^2} \left({\bar \psi^i}_L \gamma^\mu
  T^{iw}_L \right) \left({\bar T^{jw}}_L \gamma_\mu \psi^j_L \right)\ .
\label{unfierz}
\end{equation}
When Fierzed
into the form of a product of technicolor singlet currents, this includes
\begin{equation}
  -{\xi^2\over 2} {\gE^2\over
  \ME^2}\left({\bar\psi}_L\gamma^\mu\tau^a\psi_L \right) \left({\bar T}_L
  \gamma_\mu \tau^a T_L \right)\ ,
\label{fourferm}
\end{equation}
where the $\tau^a$ are weak isospin Pauli matrices.  We will show that
operator (\ref{fourferm}) can generate sizeable deviations in the
predictions  for the $\Zbb$ coupling.  The full Fierzed form of operator
(\ref{unfierz}) also includes operators involving products of weak-singlet
left-handed currents, but those will not affect the $\Zbb$ coupling.

Our analysis of operator (\ref{fourferm}) proceeds along the lines of  Ref.
10. Adopting an effective chiral Lagrangian description appropriate below
technicolor's chiral symmetry breaking scale, we may replace the
technifermion current by a sigma-model current\cite{11} :
\begin{equation}
   \left({\bar T}_L \gamma_\mu \tau^a T_L \right) =
	{v^2 \over 2}Tr\left(\Sigma^\dagger\tau^a iD_\mu\Sigma\right)\ ,
\end{equation}
where $\Sigma = \exp{(2i{\tilde\pi}/v)}$ transforms as
$\Sigma \rightarrow L\Sigma R^\dagger$ under $SU(2)_L \times SU(2)_R$,
and the covariant derivative is
\begin{equation}
  \partial_\mu\Sigma
+ i{e\over\sw\sqrt2}\left(W_\mu^+\tau^+ + W_\mu^-\tau^-\right)\Sigma
+ i\esc Z_\mu\left({\tau_3\over 2}\Sigma - \sw^2[Q,\Sigma]\right)
+ ieA_\mu[Q,\Sigma]\ .
\end{equation}
In unitary gauge $\Sigma=1$ and operator (\ref{fourferm}) becomes
\begin{equation}
   {\xi^2\over 2} {\gE^2 v^2\over\ME^2}{\bar\psi}_L \left(
 \esc Z\!\!\!\!/ {\tau_3\over 2} +
 {e\over\sw\sqrt2}\left(W\!\!\!\!\!/\ ^+ \tau^+ + W\!\!\!\!\!/\ ^-
  \tau^-\right) \right)\psi_L\ .
\label{zslashop}
\end{equation}
This yields a correction
\begin{equation}
\delta g_L = -{\xi^2 \over 2} {\gE^2 v^2\over\ME^2} \esc(I_3)
	= {\xi^2 \over 4} {m_t\over{4\pi v}} \cdot \esc
\end{equation}
to the tree-level $\Zbb$ coupling $g_L = \esc (I_3-Q\sw^2) =
 \esc(-\half + {1\over 3}\sw^2)$ and $g_R = \esc {1\over 3}\sw^2$.

Consider the effect of $\delta g_L$ on the ratio of the $b\bar{b}$ and
hadronic widths of the $Z$. In any such ratio of widths, all oblique
effects and, in particular, any effects from the $\rho$ parameter
approximately cancel.  The change in this ratio is
\begin{equation}
\delta \left( {\Gamma_\bb \over \Gamma_{had} }\right)
\approx \left( {\Gamma_\bb \over \Gamma_{had}}\right) \left(
{\delta \Gamma \over \Gamma_\bb} - {\delta \Gamma \over \Gamma_{had}}
\right),
\end{equation}
where $\delta \Gamma$ is the purely non-oblique correction to the $\Zbb$
width
\begin{equation}
  {\delta\Gamma\over\Gamma_\bb} \approx { 2{g_L\delta g_L}\over
   {g_L^2+g_R^2}} \approx -3.7\% \cdot \xi^2
   \left({m_t\over 100\ {\rm GeV}}\right)\ .
\label{dgamma}
\end{equation}
For a top mass of order 100 GeV, the standard model predictions\cite{12}
are 378 MeV for $\Gamma_\bb$ and 1734 MeV for $\Gamma_{had}$, so
that (\ref{dgamma}) leads to
\begin{equation}
\delta \left({\Gamma_\bb \over \Gamma_{had}}\right) \approx
- 2.9\% \xi^2 \left({m_t \over 100\ {\rm GeV}} \right)
\left({\Gamma_\bb \over \Gamma_{had}}\right).
\label{whynot}
\end{equation}

For comparison, we shall briefly consider the $\Zbb$ corrections in the
one-Higgs standard model and in technicolor models with extra light scalars
coupling to ordinary and techni-fermions\cite{3,4}.  The $\Zbb$ effects in
the two kinds of models are practically identical; in both models the
leading $m_t$-dependence arises from exchange of longitudinal $W^\pm$ bosons
(charged Goldstone bosons) and is {\it quadratic} rather than {\it linear}
in $m_t$.   This vertex correction alters the ratio of $\bb$ and hadronic
widths by approximately\cite{13} 0.5\% if $m_t=$100 GeV and 2.0\% if
$m_t=$200 GeV.  The technicolor models do have additional contributions at
order $m_t^2$ due to exchange of the extra charged scalars, but these are
generally negligible since they are suppressed by a factor of
$(\lambda_T f_{tc} / M_\phi)^2$ where $\lambda_T$ is the scalar coupling
to technifermions, $M_\phi$ is the scalar mass\cite{4}, and $f_{tc}$ is the
technicolor $F$ constant (which is less than $v$ in these models).

Experiments at LEP currently\cite{14,15} measure $\Gamma_\bb/\Gamma_{had}$
to an accuracy of about 5\%, so a shift on the
order of (\ref{whynot}) cannot yet be excluded.  The
measurement of $\Gamma_\bb/\Gamma_{had}$ should\cite{14} eventually
reach 2\%, at which point it will be possible to either observe or exclude
the effect in Eq.~(\ref{whynot}). We also note that the potentially large
correction to the $Wtb$ vertex arising in Eq.~(\ref{zslashop}) will be
much more difficult to observe without detailed studies of the top quark.

Finally, we consider relaxing the assumption that the ETC and weak
interactions commute. The ETC boson responsible for  generating the top
mass can still contribute to the $\Zbb$ vertex as above.  For example, the
operator (\ref{topfierz}) can arise from the exchange of a weak-doublet ETC
gauge boson which couples $T_L$ to $t^c_L$ (the field charge-conjugate to
$t_R$) and $\psi_L$ to $U^c_L$.  Additionally, such a gauge boson will give
rise to an  $SU(2)_{L+R}$ triplet operator $(\bar{U}_R \gamma^\mu U_R)
(\bar{\psi}_L \gamma_\mu \psi_L)$   which clearly affects $\Zbb$.  But
other ETC gauge bosons may also have an impact: technicolor-neutral
weak-triplet ETC bosons would contribute directly  to an operator of the
form of Eq.~(\ref{fourferm}).  We would generically expect the effects on
the $\Zbb$ coupling from either operator  to be of the same order of
magnitude as those already described, but  the size and sign of the total
shift (\ref{dgamma}) would be model-dependent. Hence, in theories with
weak-charged gauge bosons, we find that we can  make no definite
predictions.

\bigskip
We thank A. Cohen, H. Georgi, D. Kaplan, J. Kroll, K. Lane, J. Thomas and
B. Zhou for conversations and comments.  R.S.C. acknowledges the support of
an NSF Presidential Young Investigator Award and of an Alfred P. Sloan
Foundation Fellowship.  S.B.S. and E.H.S. acknowledge the support of SSC
Fellowships from the TNRLC.  Work supported in part by NSF contracts
PHY-9057173 and PHY-8714654, by DOE contracts DE-AC02-89ER40509  and
DE-FG02-91ER40676 and by TNRLC grants RGFY91B6 and RGFY9106.

\bibliographystyle{unsrt}

\end{document}